# Towards Bloodless Potassium Measurement from ECG using Neuro-Fuzzy Systems

Zeynab Samandari[1], Seyyedeh Fatemeh Molaeezadeh[2]*

[1]MSc Student, Department of Electrical and Computer Engineering, Jundi-Shapur University of Technology, Dezful, Iran. (https://orcid.org/0000-0003-1680-3636)

[2]Assistant Professor, Department of Electrical and Computer Engineering, Jundi-Shapur University of Technology, Dezful, Iran. (https://orcid.org/0000-0003-0921-6310)

* Corresponding author:

Phone: (+98) 061-42418500-2346

Fax: (+98) 061-42418555

Email: fmolaee@jsu.ac.ir

**Abstract**

Potassium disorders are generally asymptomatic, potentially lethal, and common in patients with renal or cardiac disease. The morphology of the electrocardiogram (ECG) signal is very sensitive to the changes in potassium ions, so ECG has a high potential for detecting dyskalemias before laboratory results. In this regard, this paper introduces a new system for ECG-based potassium measurement. The proposed system consists of three main steps. First, cohort selection & data labeling were carried out by using a 5-minute interval between ECGs and potassium measurements and defining three labels: hypokalemia, normal, and hyperkalemia. After that, feature extraction & selection were performed. The extracted features are RR interval, PR interval, QRS duration, QT interval, QTc interval, P axis, QRS axis, T axis, and ACCI. Kruskal-Wallis technique was also used to assess the importance of the features and to select discriminative ones. Finally, an ANFIS model based on FCM clustering (FCM-ANFIS) was designed based on the selected features. The used database is ECG-ViEW II.  Results showed that T axis compared with other features has a significant relationship with potassium levels (P<0.01, r=0.62). The absolute error of FCM-ANFIS is 0.4±0.3 mM, its mean absolute percentage error (MAPE) is 9.99%, and its r-squared value is 0.74. Its classification accuracy is 85.71%. In detecting hypokalemia and hyperkalemia, the sensitivities are 60% and 80%, respectively, and the specificities are 100% and 97.3%, respectively. This research has shed light on the design of noninvasive instruments to measure potassium concentration and to detect dyskalemias, thereby reducing cardiac events.

# 1. Introduction

Potassium is of critical importance for the normal functioning of excitable cells such as nerve and muscle ones as well as cardiac rhythm regulation, so that potassium concentration is tightly regulated by homeostatic mechanisms [1-3]. Fluctuations in potassium concentrations even on modest amounts could increase a potential risk of morbidity, hospitalization, and mortality in renal or cardiovascular patients [3-6]. Dyskalemias -hyperkalemia and hypokalemia- are often clinically silent and may lead to life-threatening arrhythmias and neuromuscular dysfunction [2, 3, 5, 7, 8].

There are a couple of varied factors leading to potassium disorders. They may occur as a result of medication use, hyperglycemia, gastrointestinal losses, hypertension, aging, and potassium-rich food intake [2-5, 9]. Retrospective studies revealed that hypokalemia and hyperkalemia are common in patients with heart failure and renal dysfunction, respectively [1, 2, 4, 9, 10]. Some drugs such as diuretics have the potential to cause hypokalemia [2] whereas other medications such as non-steroidal anti-inflammatory drugs can lead to hyperkalemia [5]. It is ironic that heart failure medications (e.g., aldosterone antagonists) used to treat hypokalemia increase mortality rate due to hyperkalemia [3-5]. Therefore, monitoring potassium levels is absolutely vital in order to minimize medicine risks and to administrate proper drug therapies. Moreover, due to the crucial importance of potassium homeostasis for hemodialysis patients, it can be of extreme help in designing personalized hemodialysis sessions tailored to the patient's specific needs [10].

Currently, potassium abnormalities are identified with diagnostic laboratory tests [2, 4, 7, 8, 11]. The expected turnaround time to assess potassium for critically ill patients is within a couple of minutes [12]. Blood samples poorly collected or destroyed during transportation can lead to erroneous laboratory test results and repeat testing, thereby making a delay in potassium management. Commonly, hospitalized patients, especially elderly ones, have difficult veins and face the potential risk of anemia as a consequence of repeat blood draws [13, 14]. Beyond higher costs and required trained technicians, phlebotomy is difficult and painful for high-risk patients [7, 14]. Therefore, a continuous, convenient, cost-effective and bloodless method for potassium monitoring is highly desirable for promptly recognizing and timely treating dyskalemias [2-5, 7, 8, 15-17].

The electrocardiogram (ECG) is a non-invasive, patient-friendly, quick and low-cost technique that reflects the heart's electrical activity. It is well known that potassium abnormalities have profound impacts on cardiac conduction and thus ECG [4, 6, 8-10], [18-20]. For instance, electrocardiographic findings show that hypokalaemia may produce flattened or inverted T waves, a ST segment depression, a prominent U wave, and a prolonged QT interval [5, 8, 19, 20] while hyperkalaemia may make tall peaked or tented T waves, PR interval prolongation, loss of P wave, and QRS complex widening [5, 8, 9, 19, 20]. Therefore, electrocardiographic changes may provide useful clues to diagnosis potassium abnormalities before laboratory results preparation.

To the best of our knowledge, modelling studies in this context are relatively at the basic level, mainly focus on either regression or classification to automatically detect an impaired potassium homeostasis. Some studies have been just concentrated on a quantification of T-wave morphology and a correlation analysis between potassium levels and T-wave-derived features [5, 18]. Models developed to estimate potassium concentrations are the quadratic regression model [10], linear mixed model [4, 9], linear regression model [3, 7, 21], and 82-layer Convolutional Neural Network (CNN) [8]. Hyperkalemia detection has been done based on methods such as a two-stage K-means classifier [17], two-stage artificial neural network [16], and an 11-layer CNN [22]. An 82-layer CNN is also used for classifying dyskalemias [8].

Fuzzy systems attempt to mimic human knowledge and reasoning processes in an approximate manner [23, 24]. These systems are expert-driven and glass-box models as well as being properly suited for dealing with ill-defined and uncertain systems [23, 24]. They can be modelled to be highly interpretable and transparent for users [23, 25]. However, the accuracy of these models when faced with highly complex problems is criticized [25, 26]. On the other hand, neural networks represent the brain's architecture in a precise manner [23]. These systems are data-driven and opaque-box models meanwhile they have high accuracy solving more sophisticated problems owing to learning capability [26]. Nevertheless, due to the huge number of layers and features especially in deep neural networks, these systems suffer from non-transparency and might not be interpretable for humans [15, 23, 25, 26]. Neuro-fuzzy system is a hybrid architecture that possesses the features of neural networks accuracy and the fuzzy systems transparency, simultaneously. Hence, these systems have attracted researchers' interests especially in the realm of explainable artificial intelligence (XAI) [23, 25, 26, 27] and could hold possible promise for constructing XAI systems [23]. This potential prompted us to develop a neuro-fuzzy model to assist physicians in diagnosing and treating dyskalemias.

Our contributions in this paper are as follows. We firstly assessed the correlation of the age-adjusted Charlson comorbidity index (ACCI) and eight ECG-derived features with potassium levels. After that, we designed two adaptive-network-based fuzzy inference systems (ANFIS): Grid-partitioning-based ANFIS (conventional ANFIS) and FCM-clustering-based ANFIS (FCM-ANFIS) to estimate serum potassium level then we compared the performance of both of them in terms of estimation error. Then, we evaluated the diagnostic performance of the proposed model in detecting both hypokalemia and hyperkalemia. Finally, we summarized the results of previous works as a helpful review of the noninvasive potassium measurement field.

This paper is organized as follows. Section II describes the proposed framework for automatic potassium quantification and classification. Section III presents the results in terms of recognizing statistically relevant features with potassium level and evaluating the performance of the conventional ANFIS and FCM-ANFIS models for estimating potassium concentration and detecting potassium disorders. Section IV discusses the results and compares them with the results of other studies. Finally, section V summarizes the results and findings of this study.

## 2. Materials and Methods

Fig.1 provides a brief overview of the proposed methodology in this study. Our method includes four steps: cohort selection & data labelling, feature extraction & selection, fuzzy modelling, and model evaluation. ECG-ViEW II database is used in this study. In the following, details of each step are provided.

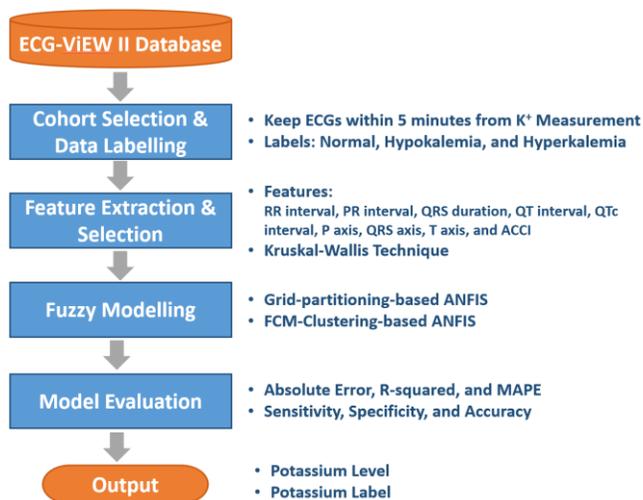

**Fig. 1.** Block diagram of the proposed methodology for the ECG-based estimation of potassium concentration.

### 2.1 ECG-ViEW II Database

ECG-ViEW II Database is a freely accessible, large, single-center, and real-world database including 979,273 electrocardiograms from 461,178 patients over a 19-year study period. The average and the standard deviation of patients' age are 42.6 and 19.2 years, respectively. This database contains numeric ECG parameters (RR interval, PR interval, QRS duration, QT interval, QTc interval, P axis, QRS axis, and T axis), demographic data (age, gender, and ethnicity), and clinical data (medications, diagnoses, and laboratory tests). ACCI was also calculated when ECG was performed. This database can offer an opportunity to study the relationship between electrolyte concentration and ECG [28].

**2.2 Cohort Selection & Data Labelling**

In ECG-ViEW II Database, there are potassium concentrations for only a number of patients (n=34,384 patients) and some patients had also all ECG parameters at the same time (7,864 patients). In this case, for patients having several ECG parameters for a specific potassium measurement, we keep ECG parameters with closest time to the potassium measurement. 42 out of 461,178 patients were finally enrolled who had simultaneously all ECG parameters and ACCI within 5 minutes before or after potassium measurements. These selected patients aged 18-90 years and had different diseases such as cerebral haemorrhage, pneumonia, supraventricular tachycardia, hematochezia, septicemia and dilated cardiomyopathy. Fig.2 briefly overviews the cohort selection process.

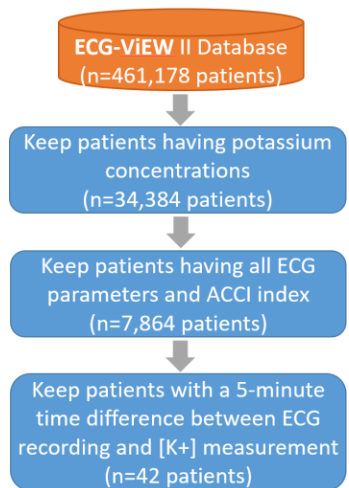

**Fig. 2.** Cohort selection process from ECG-ViEW II database to the dataset for model training.

Potassium levels were classified into three labels: hypokalemia (<3.5 mM), normal (3.5-5 mM), and hyperkalemia (> 5 mM) [29]. In the selected dataset, there are 10, 27, and 5 patients with hypokalemia, normal potassium, and hyperkalemia, respectively.

*C. Feature Extraction & Selection*

Since ECG time series does not exist in the database, this paper applies nine ECG parameters mentioned in subsection II.A as ECG-extracted features. After that, Kruskal–Wallis one-way analysis of variance (ANOVA) were carried out to evaluate the significance of these selected features in differentiating three labels: hypokalemia, normal, and hyperkalemia. In this way, the key features were selected to enter into the proposed model. Pearson correlation coefficient was computed for the selected features. In our statistical analyses, the p-value <0.05 was considered significant.

**2.3 Fuzzy Modelling & Model Evaluation**

To estimate potassium level, the FCM-ANFIS model was applied and the performance of the proposed model was also compared with the conventional ANFIS. The estimation performance was evaluated by two criteria. The first criterion is the correlation coefficient between the estimated values and the actual values and the second one is the mean absolute percentage error (MAPE) computed as follows:

$$MAPE = \frac{1}{n}\sum_{i=1}^{n}\left|\frac{\hat{y}_i - y_i}{y_i}\right| \times 100 \qquad (1)$$

, in which $\hat{y}_i$ is the estimated value; $y_i$ is the actual value; and $n$ is the number of data points.

We classified the estimated values obtained the model into three labels and evaluated the diagnostic performance of the proposed model for these labels. In this case, we use confusion matrix and its extracted criteria such as sensitivity, accuracy, and specificity to evaluate the model performance. 10-fold cross validation (10-fold CV) was applied to split data into two categories: train and test.

## 3. Results

Fig.3 draws a comparison of the Chi-square values obtained from Kruskal-Wallis test in all mentioned features. As it can be seen, relevant features are T axis (Chi-square value=15.61, p value=0.0004), QTc interval (Chi-square value=7.87, p value=0.019), and ACCI (Chi-square value=7.13, p value=0.028). Therefore, T axis is statistically the most significant. The correlation coefficient between T axis and potassium levels is 0.62.

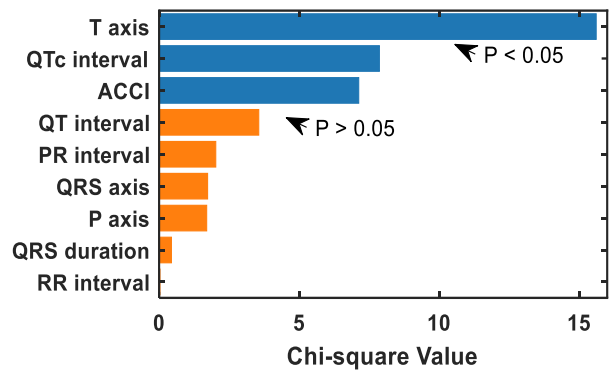

**Fig. 3.** A comparison of the Chi-square values obtained from Kruskal-Wallis test in RR interval, PR interval, QRS duration, QT interval, QTc interval, P axis, QRS axis, T axis, and ACCI.

In order to have a better insight of discrimination power among three labels (hypokalemia, normal potassium, and hyperkalemia) of the three selected features, we illustrated boxplot charts for them in Fig.4. As it can be seen, the interquartile range (IQR) with a 95% confidence interval (CI) reveals a proper distinction between three labels for T axis feature. As a result, this feature is applied as an input to the fuzzy model for estimating potassium levels.

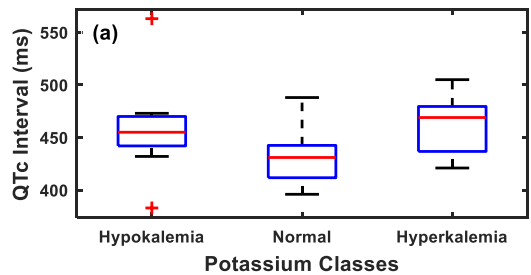

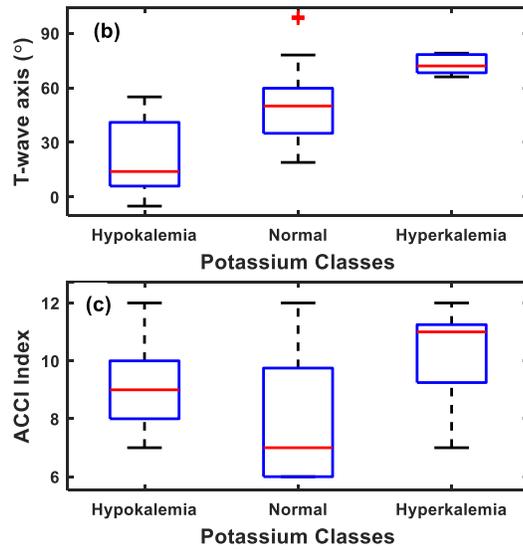
Fig. 4. Boxplot chart showing a comparison of (a) QT interval, (b) T-wave axis, and (c) ACCI index for three groups: hypokalemia, normal, and hyperkalemia.

By a try-and-error method, we chose a linear function for output. In other words, first-order ANFIS models were applied in this study. We considered five trapezoidal membership functions for the conventional ANFIS and three clusters for FCM-ANFIS. Fuzziness parameter and stopping criterion in FCM clustering method were set to 2 and 0.00001, respectively. Training the models were also run 200 epochs. Consequent parameters were adjusted by the least mean square error (LMS) algorithm in both models. Antecedent parameters in the conventional ANFIS were trained by Gradient descent algorithm. For FCM-ANFIS, we firstly set clusters by FCM clustering algorithm during the first 100 epochs. Then, we fine-tuned membership function parameters for the clusters during the last 100 epochs. Furthermore, 10-fold CV is used for determining training and test dataset. Hence, ten models are trained for each of ANFIS systems.

Fig.5 shows the results of these ten models for both ANFIS in boxplot charts. As it can be seen, the IQR with a 95% CI in FCM-ANFIS is overlapped with target data (actual potassium concentration) to a greater extent in comparison with the conventional ANFIS.

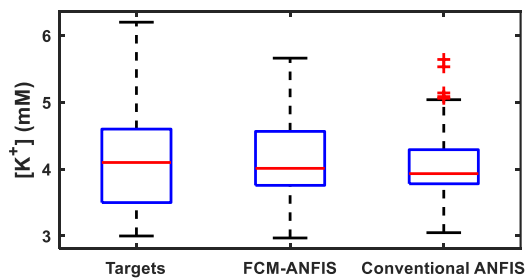
Fig. 5. Boxplot charts showing a comparison of the $[K^+]$ estimation performance between FCM-ANFIS and conventional ANFIS.

Table 1 shows a quantitative comparison between conventional ANFIS and FCM-ANFIS for potassium concentration estimation. Results shows that FCM-ANFIS has a higher performance in terms of less error, absolute error and MAPE. The correlation coefficient between the FCM-ANFIS estimated $[K^+]$ and the target data (the actual $[K^+]$) is 0.74.

TABLE I
A quantitative comparison of the $[K^+]$ estimation performance between FCM-ANFIS and conventional ANFIS

| Model | Error (mM) | Absolute error (mM) | MAPE (%) |
| --- | --- | --- | --- |
| Conventional ANFIS | -0.006±0.52 | 0.42±0.31 | 10.34 |
| FCM-ANFIS | -0.01±0.50 | 0.40±0.30 | 9.99 |

The generated rules and membership functions in FCM-ANFIS are shown in Table 2 and Fig.6. As it is illustrated, the model is explainable in terms of three rules concluding that T axis is directly proportional to [K$^+$].

TABLE II
The final rules generated in FCM-ANFIS

| No. | Rules |
|---|---|
| 1 | If T axis is Low then [K$^+$] = -0.0501×T axis + 6.9810 |
| 2 | If T axis is Medium then [K$^+$] = -0.0712×T axis + 8.0007 |
| 3 | If T axis is High then [K$^+$] = -0.1123×T axis + 8.9554 |

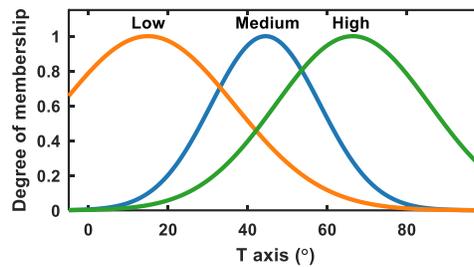

Fig. 6. The final antecedent membership functions for FCM-ANFIS

Fig.7 provides a comparison between FCM-ANFIS and conventional ANFIS in view point of dyskalemias detection. According to the confusion matrix in Fig. 7, FCM-ANFIS classification accuracy is 85.71%. Moreover, in detecting hypokalemia and hyperkalemia, the sensitivities were 60% and 80%, respectively. In addition, the specificities were 100% and 97.3%, respectively. Conventional ANFIS classification accuracy is 80.95 %. In detecting hypokalemia and hyperkalemia, also the sensitivities were 50% and 80%, respectively. Furthermore, the specificities were 100% and 94.59%, respectively.

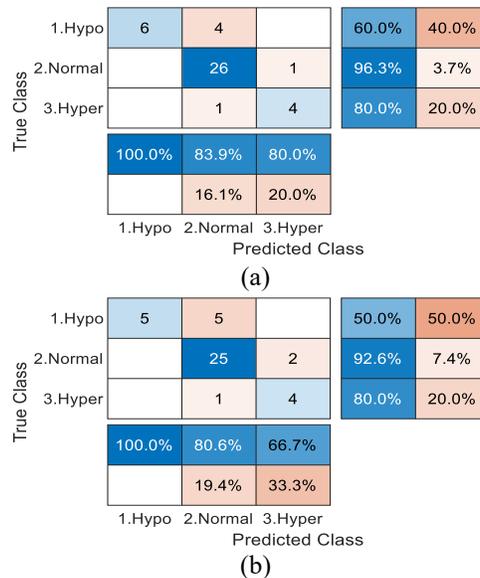

Fig. 7. Confusion matrixes showing a comparison of the dyskalemias detection performance between (a) FCM-ANFIS and (b) conventional ANFIS.

## 4. Discussion

This study firstly verified the importance of ECG-derived features and ACCI in [K$^+$] estimation. The results showed that T axis is a dominant feature. Rightward T axis plays a significant role in hyperkalemia while leftward T axis has a similar role in hypokalemia. These findings were in agreement with the results of previous studies. T axis deviation has been reported as a marker of ventricular repolarization abnormalities and a potential indicator of increased risk for cardiovascular mortality [28], [30].

Moreover, T axis was a significant variable in hyperkalemia as well as renal impairment. In moderate-to-severe renal impairment, ECG signal had a rightward T axis [8, 31].

After that, this study designed a neuro-fuzzy system for [K$^+$] estimation based on T axis and applied it for dyskalemias detection. Tables 3&4 compare the performance of the proposed system with other works in view point of regression and classification tasks, respectively. As it can be seen, our proposed system has promising results.

In our study, [K$^+$] estimation was based on only 5 minutes of ECG before or after blood draw for potassium measurement. As the [K$^+$] is dynamically slow [1], [K$^+$] estimation less than five minutes is improbable to add clinical information. Furthermore, a 5-minute timeframe may be more applicable in practical cases for continuous or remote monitoring situations [3]. Previous

Table 3. A comparison of our work with previous studies in view of regression task

| Work/Year | Patients (Type, No.) | Time frame | No. of Leads | K$^+$ Range (mM) | Features | Model | Type of Model | MAPE (%) | Absolute Error (mM) | Error (mM) |
|---|---|---|---|---|---|---|---|---|---|---|
| Attia et. al. [3], 2016 | Hemodialysis, 26 | 5 min | 4 | 3.9±0.8 | $T_{S/\sqrt{A}}$ | First-order polynomial | Personalized | 10 | 0.36±0.34 | N/A |
|  |  |  |  |  |  |  | Global | 11 | 0.44±0.47 |  |
|  | Hemodialysis, 19 |  |  | 4.2±0.95 |  |  |  | 12 | 0.5±0.42 |  |
| Yasin et. al. [7], 2017 | Hemodialysis, 18 | 2 min | 1 | 4.3±0.8 | $T_{S/\sqrt{A}}$ | First-order polynomial | Personalized | 9 | 0.38±0.32 | N/A |
| Corsi et. al. [10], 2017 | Hemodialysis, 45 | 15 min | 12, PCA | 2.5 to 7.5 | $T_{S/A}$ | Second-order polynomial | Personalized | N/A | 0.46±0.39 | -0.09±0.59 |
|  | LQT2, 12 |  |  | 3.9 to 4.8 |  |  |  |  | N/A | 1.1±1.3 |
| Lin et. al. [8], 2020 | Emergency, 40180 | ±1hr | 12 | 1.5 to 7.5 | CNN | CNN | Global | N/A | 0.531 | N/A |
| **This work** | Different, 42 | ±5 min | 2 | 3 to 6.2 | T axis | FCM-ANFIS | Global | **9.99** | **0.40±0.30** | **-0.01±0.50** |

$T_{S/A}$: T-wave slope-to-amplitude ratio; $T_{S/\sqrt{A}} = T\_right\_slope / \sqrt{T\_amplitude}$; mM: mmol/L; PCA: Principal Component Analysis; LQT2: Long QT syndrome type 2.

Table 4. A comparison of our work with previous studies in view of classification task

| Work/Year | Patients (Type, No.) | Time frame | No. of Leads | Features | Model | Class | Sen (%) | Spe (%) | Acc (%) | ROC-AUC |
|---|---|---|---|---|---|---|---|---|---|---|
| Wu et. al. [16], 2003 | Emergency, 50 | N/A | 12 | T wave amplitude and duration P wave amplitude and duration QRS duration, PR interval averaged RR interval | ANN | [K$^+$]↑ | 60 | 65 | 62.5 | N/A |
| Tzeng et. al. [17], 2005 | Emergency, 97 | N/A | 12 | T-wave volume, PR interval, QRS duration, QT interval | K-means | [K$^+$]↑ | 85 | 79 | N/A | N/A |
| Velagapudi et. al. [9], 2016 | Hyperkalemia, 107 | ±4hr | 12 | T width, T descending slope, QRS width | Linear mixed model | [K$^+$]↑ | 63 | 84 | N/A | 0.78 |
| Galloway et. al. [22], 2019 | CKD, 511345 M | 4 hr | 2 | CNN | CNN | [K$^+$]↑ | 90.2 | 63.2 | N/A | 0.883 |
|  | F |  |  |  |  |  | 91.3 | 54.7 |  | 0.86 |
|  | A |  |  |  |  |  | 88.9 | 55 |  | 0.853 |
| Lin et. al. [8], 2020 | Emergency, 40180 | ±1hr | 12 | CNN | CNN | [K$^+$]↓ | 67.5 | 93.3 | N/A | N/A |
|  |  |  |  |  |  | [K$^+$]↑ | 67.5 | 97.8 |  |  |
| This work | Different, 42 | ±5 min | 2 | T axis | FCM-ANFIS | [K$^+$]↓ | 60 | 100 | 85.71 | N/A |
|  |  |  |  |  |  | [K$^+$]↑ | 80 | 97.3 |  |  |

Sen: Sensitivity; Spe: Specificity; Acc: Accuracy; N/A: Not Available; ANN: Artificial Neural Network; Time frame is a time interval between ECG recordings and [K+] measurements; M: Minnesota; F: Florida; A: Arizona.

studies have considered the time windows of 2 minutes [7], 5 minutes [3, 5], 15 minutes [10], 1 hour [8], and 4 hours [9, 22]. To recapitulate, time delays within a couple of minutes or hours to determine a potassium value are generally acceptable in clinical practices [3].

The proposed neuro-fuzzy system was designed in a global manner and did not require any patient-specific calibration. This system is a shallow and explainable model on a visual platform in the format of if-then rules. As a way forward to XAI, Lin *et. al.* [8] tried to open the back box of 82-layer CNN using a visualization approach. Galloway *et. al.* used an 11-layer CNN. This paper provides a three-class classification and evaluates both hypo- and hyper-kalemia.

The main limitation of this paper is the lack of benchmark and public dataset relating ECG changes to potassium levels. To the best of our knowledge, ECG ViEW II is the only public dataset in this area. This dataset is composed of information extracted from 461,178 patients over a 19-year study period. However, this dataset is faced with a variety of problems. Firstly, it does not contain any ECG time series. Hence, this study is limited to the nine features mentioned in this dataset and could not utilize from other ECG-derived parameters especially T wave-based morphological features such as T-wave slope, T-wave amplitude, and T-wave slope-to-amplitude ratio. Secondly, there are different timeframes between [$K^+$] and ECG-based features. Therefore, despite of the high number of data in this dataset, this study is limited to only 42 patients with a 5-minute time window. Lastly, the majority of the data is related to normal potassium concentrations and this is the main reason for unbalanced classes, thereby increasing a learning rate in the normal class.

Another limitation is the erroneous laboratory results happened as a result of the unavoidable delay associated with transport to the clinical laboratory and exposure to changing temperature of ambient [32]. The occurrence of spurious hypo- and hyperkalemia is common in summer and winter months, respectively [32]. Pseudohyperkalemia may relate to normal electrocardiogram or it can mask hypokalemia by pushing measured values into the normal range [12]. These incorrect results can be a great source of diagnostic error and they potentially expose patients to danger [32]. Hence, it is necessary to evaluate these cases by the laboratory or the clinician and to remove them from dataset.

Last limitation is related to reliable automatic extraction of ECG waves. ECG is very prone to background noises from skeletal muscle or skin galvanic current as well as confounding factors such as heart rate, body position, and other metabolic or cardiovascular conditions, resulting in a poor signal-to-noise ratio [4]. Hence, it hampers the detection of morphological features such as QRS complex which is the first step in all kinds of automated feature. Furthermore, morphologies of many normal QRS complexes and abnormal ones differ widely as well as other ECG waves (such as P and T waves) can hinder QRS complex detection [33]. T wave detection plays a significant role in quantifying potassium concentrations and a poor SNR makes T wave detection challenging. T wave also overlaps with P wave and it has a low amplitude, energy, and frequency so it is prone to get corrupted by noise to a high extent. Moreover, as another challenge, T wave has great variations in its morphologies such as peaked, inverted, flat, positive, negative, biphasic, and camel hump shaped T waves [34-37].

## 5. Conclusion

In the estimation of [$K^+$], T axis had a direct and statistically significant correlation with [$K^+$]. Moreover, our work demonstrates the promising application of FCM-ANFIS as a XAI model in the field of potassium estimation and dyskalemias detection. FCM-ANFIS could detect hypo- and hype-Kalemia with an acceptable and proper sensitivity and a high specificity. Besides, in comparison with conventional ANFIS, the proposed system had improved the MAPE and classification accuracy by 3.5 % and 5.88%, respectively. This system was based on only 5 minutes of ECG which may be more practical and helpful for alerts and trending in continuous monitoring applications. This study may be a subject for further investigations in large cohorts of patients.

## Declarations

### Ethical Approval
Not applicable.

### Competing interests
Authors have no conflict of interest.

### Authors' contributions
Z.S. implemented the proposed Methodology under supervision F. M. and wrote the initial draft of manuscript. F. M wrote the final draft of manuscript and edited figures. All authors reviewed the manuscript.

**Funding**
Not applicable.

**Availability of data and materials**
ECG-ViEW II Database is publically available.